\documentclass[%
 aip,
rsi,%
 amsmath,amssymb,
 reprint,%
]{revtex4-1}

\usepackage{graphicx}% Include figure files
\usepackage{dcolumn}% Align table columns on decimal point
\usepackage{bm}% bold math
\begin{document}

\preprint{AIP/123-QED}

\title{Structural, magnetic, thermodynamic and electrical transport properties of a new compound $\mbox{Pr}_2\mbox{Rh}_{2}\mbox{Ga}$}

 \author{Baidyanath Sahu}
  \altaffiliation{Corresponding author}%Lines break automatically or can be forced with \\
  
  \author{Sindisiwe P. Xhakaza}
   \affiliation{Highly Correlated Matter Research Group, Physics Department, University of Johannesburg, PO Box 524, Auckland Park 2006, South Africa%\\This line break forced% with \\
  }%
  
 \author{Andr\'{e} M. Strydom}%
  %\email{Second.Author@institution.edu.}
 \affiliation{Highly Correlated Matter Research Group, Physics Department, University of Johannesburg, PO Box 524, Auckland Park 2006, South Africa%\\This line break forced with \textbackslash\textbackslash
 }%


\begin{abstract}
A new ternary intermetallic compound $\mathrm{Pr_2Rh_2Ga}$ was synthesized by arc-melting and was  characterized by powder X-ray diffraction (PXRD), magnetization, heat capacity $\mathrm{C}_p(\textit{T})$, and electrical resistivity $\rho(T)$ measurements. PXRD patterns revealed that $\mathrm{Pr_2Rh_2Ga}$ crystallizes in the $\rm{La_2Ni_3}$-type of orthorhombic structure with the space group $Cmca$. The temperature variation of magnetic susceptibility, $\mathrm {C}_p(\textit{T})$ and $\rho(T)$ confirmed that $\mathrm{Pr_2Rh_2Ga}$ exhibits a ferromagnetic behavior with the transition temperature of 18 K. The estimated Sommerfeld coefficient $\gamma$ = 640 mJ/($\mathrm{Pr.mole.K^2}$) from the $\mathrm {C}_p(\textit{T})$ results in the paramagnetic region just above $T_{C}$~was large in comparison to ordinary metals. In the paramagnetic region of $\rho(T)$ data showed a metallic behavior characteristic of electron - phonon scattering. The maximum negative magneto-resistance at high field occurs in the region near the magnetic phase transition temperature.  The maximum value of magnetic entropy change ($\rm -\Delta \textit{S}_{M}$) and adiabatic temperature change ($\rm \Delta \textit{T}_{ad}$) are $\rm8.2~J/kg.K$ and $\rm3.6~K$, respectively, around the transition temperature for the change of magnetic field $0\textendash\rm 9~T$. The calculated refrigerant capacity is $\rm70~J/kg$,  and $\rm135~J/kg$ for a change of magnetic field $0\textendash\rm 5~ T$ and $0\textendash\rm 9~ T$, respectively. Arrott plot derived from isothermal magnetization and the universal scaling plot by normalizing $\rm -\Delta \textit{S}_{M}$ confirm that the compound undergoes a second order ferromagnetic to paramagnetic phase transition.
%
%Valid PACS numbers may be entered using the \verb+\pacs{#1}+ command.
 \end{abstract}

%                              % Classification Scheme.
 %Use showkeys class option if keyword
%                               %display desired
\maketitle
% \begin{quotation}
% The ``lead paragraph'' is encapsulated with the \LaTeX\ 
% \verb+quotation+ environment and is formatted as a single paragraph before the first section heading. 
% (The \verb+quotation+ environment reverts to its usual meaning after the first sectioning command.) 
% Note that numbered references are allowed in the lead paragraph.
% %
% The lead paragraph will only be found in an article being prepared for the journal \textit{Chaos}.
% \end{quotation}

%\section{\label{sec:level1}First-level heading:\protect\\ The line break was forced \lowercase{via} \textbackslash\textbackslash}
\subsection{\label{sec:level2}Introduction}

The Praseodymium (Pr) based ternary compounds present interesting magnetic, electric and thermal transport properties such as non$\textendash$magnetic ordering, ferro and antiferromagnetic behavior, heavy fermion \cite{Pr3Rh4Sn13,PrCu2In,R2Pd2In,Pr2Pd2In,Goremychkin,Pr2Rh3Ge}, Kondo behavior \cite{PrV2Al20} and superconductivity \cite{PrOs2Pb12}. The magnetic and transport properties of Pr-compounds are strongly influenced by the crystal electric field (CEF). Pr-based ternary compounds are also attractive for the heavy fermion superconductivity properties. Bauer $et.~al.$ \cite{PrOs2Pb12}, observed the occurrence of a heavy-fermion superconductivity state in the $\mathrm{PrOs_{2}Pb_{12}}$ compound. Similarly, Zhang $et.~al.$ \cite{PrPt4Ge12}, have reported multiband superconductivity in $\mathrm{PrPt_{2}Ge_{12}}$ compound with critical temperature of 8 K. Additionally, Pr$\textendash$compounds such as  $\mathrm{PrCu_{2}In}$ \cite{PrCu2In}, $\mathrm{PrCo_{2}B_{2}C}$ \cite{PrCo2B2C}, and $\mathrm{PrNi_{2}B_{2}C}$ \cite{PrNi2B2C} show large Sommerfeld coefficient values. 

Recently $\mathrm{RE_2T_2X}$ (RE = rare earth metal, T = transitions metal and and X = p-block elements) series of compounds are of interest for attractive structural and physical properties. $\mathrm{RE_2T_2X}$ compounds are known to form in a small number different structure type, which depends on the composition. Most of $\mathrm{RE_2T_2X}$ compounds crystallize in $\mathrm{Mo_2FeB_2}$ $\textendash$ type of tetragonal structure with space group $P_4/mbm$ \cite{R2T2X,U3Si2,Crystalbook}. A small number of compounds form a superstructure of $\mathrm{Mo_2FeB_2}$ and crystallize in $\mathrm{U_2Pt_2Sn}$-type of tetragonal structure \cite{Yb2Pt2Pb}. Some of the compounds crystallize in $\mathrm{W_2CoB_2}$ and $\mathrm{Mn_2B_2Al}$ $\textendash$ type of orthorhombic structure with space group $Immm$ and $Cmmm$, respectively \cite{R2T2X,U3Si2,Crystalbook}. Some $\mathrm{RE_2T_2X}$ compounds show structural transformation, which depends on the rare-earth size, external pressure and also on processed temperature. For example, a tetragonal to orthorhombic is found in $\mathrm{RE_2Ni_2Sn}$ compounds by varying rare-earth elements and external pressure on a particular sample. $\mathrm{Pr_2T_2X}$ (T = Cu, Ni and Pd, X = In, Sn) crystallize in the tetragonal $\mathrm{Mo_2FeB_2}$ $-$ type structure \cite{Pr2Cu2In, R2T2X, Pr2Pd2In, Pr2Ni2Sn}. $\mathrm{Pr_2Ni_2Ga}$, $\mathrm{Pr_2Ni_2Al}$ and $\mathrm{Pr_2Co_2Al}$ crystallize in $\mathrm{W_2CoB_2}$ $\textendash$ type of orthorhombic structure with space group $Immm$. $\mathrm{Pr_2Co_2Al}$ shows a structural transformation from orthorhombic to monoclinic at higher temperatures \cite{Pr2Ni2Ga, Pr2Co2Al}.  

Very recently, a new variant of the 2:2:1 formula type was reported, namely  $\mathrm{Ce_2Rh_2Ga}$ which crystallizes in monoclinic (space group $C2/c$) for without annealing and shows an ordered version of the orthorhombic $\mathrm{La_2Ni_3}$ $\textendash$ structure type with space group $Cmca$ (no-64) for annealing at higher temperature. Surprisingly, this orthorhombic compound was found to exhibit a phase transition at $\approx$125 K that is unusually high among Ce compounds \cite{Ce2Rh2Ga}. In this paper, we have investigated whether other rare-earth elements are amenable to the new 2:2:1 type structure, and report on the synthesis, structure, and physical properties of a new compound $\mathrm{Pr_{2}Rh_{2}Ga}$. The magnetocaloric effect (MCE) of this compound has also been studied from isothermal magnetization and heat capacity measurements.

\subsection{\label{sec:level2}Experimental detail}
A stoichiometric mixture of the elements Pr (99.85 wt.\% purity) Rh (99.85 wt.\% purity), and Ga (99.999 wt.\% purity) with the ratio Pr:Rh:Ga = 2:2:1 and total mass of 1 g for this compound was arc-melted on a water-cooled copper hearth using an  Edmund B\"uhler GmbH MAM-1 commercial arc furnace. The sample was melted under high purity argon (Ar) atmosphere, where the gas bottle was connected to a Monotorr high-temperature gas getterer. The sample was melted several times to ensure homogeneity. The weight losses after the melting process were confirmed to be less than 1.0 wt. \%. The as-cast sample was wrapped in tantalum foil and annealed for one week at 1073 K in an evacuated silica ampoule. Finally the sample was quenched in cold water. 

In order to study the phase purity of the annealed sample $\mathrm{Pr_{2}Rh_{2}Ga}$ was characterized by powder X-ray diffraction (PXRD) using a Rigaku diffractometer with Cu-$\mathrm{K_{\alpha}}$ radiation. Magnetic, heat capacity (${C_P}$) and electrical transport ($\rho$) properties measurements were performed on a Quantum Design commercial Dynacool, Physical Property Measurement System (PPMS) in the temperature range of 1.8 - 300 K, with applied magnetic field up to 9 T. Heat capacity was measured by employing the two-$\tau$ relaxation method. The sample is stacked and thermally coupled to the sample platform using Apiezon N grease. Electrical resistivity measurements were performed by a standard four probe contact method and using an ac-current excitation.

\subsection{\label{sec:level2}Results and Discussions}
\subsubsection{X-ray diffraction}

\begin{figure}
      \centering
%      \begin{tabular}{c}
%        %example1\\
       \includegraphics[width =3.3 in, height =2.5 in]{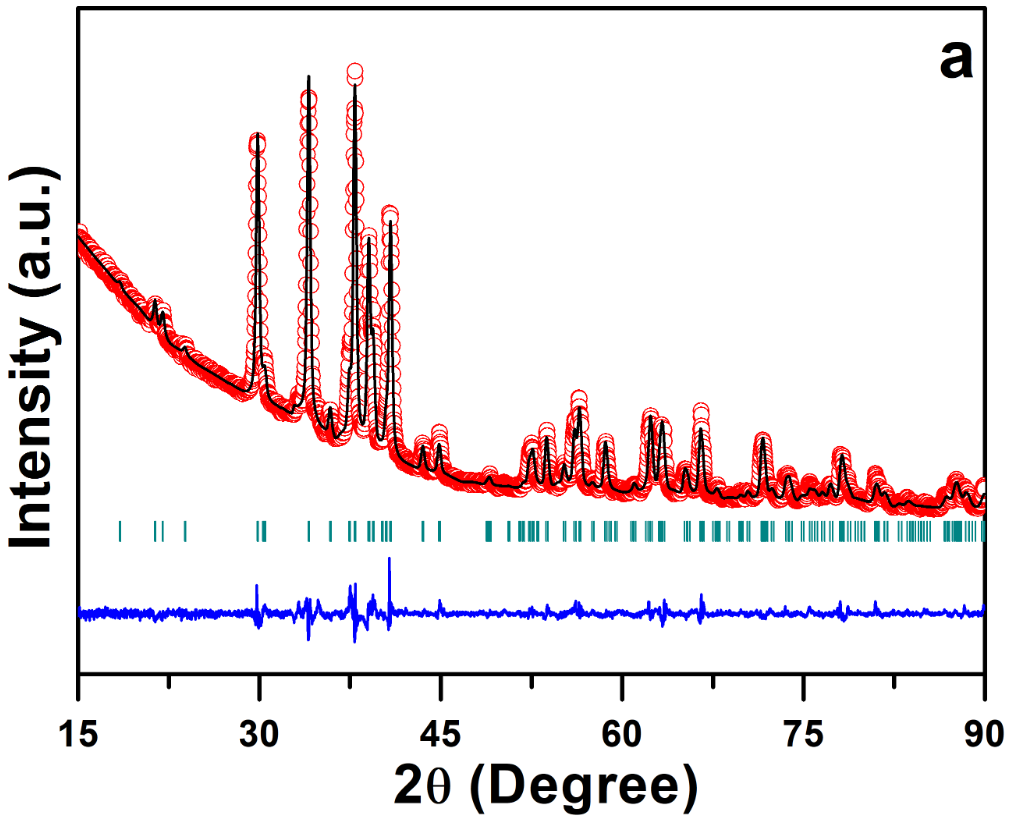}
%      \end{tabular}

      \vspace{0.03em}
        \begin{tabular}{c}
        %example2  & example3 \\
        \includegraphics[width =3.5 in, height =1.5 in]{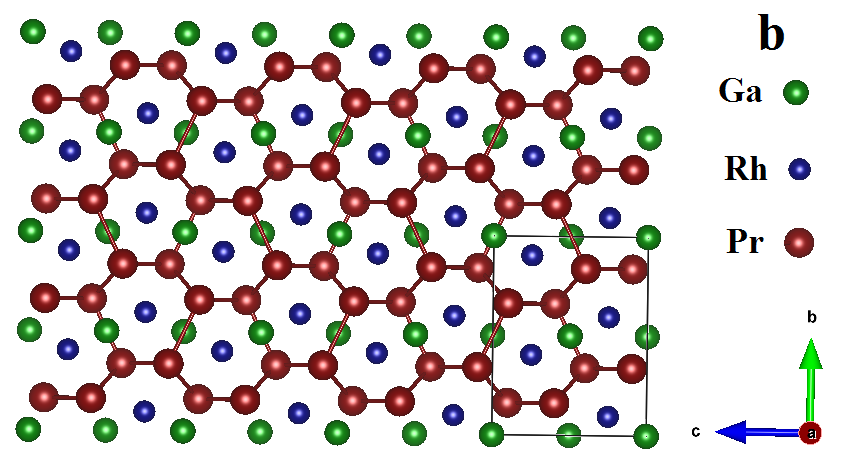}
        \end{tabular}
         \caption{(a) PXRD patterns along with the Rietveld refinement profile for $\mathrm{Pr_{2}Rh_{2}Ga}$. The red open circles represent the experimental data and black solid line stands for the calculated pattern from the model structure used in the Rietveld refinement to fit the experimental data. The difference curve is	shown as a blue line and the allowed Bragg peaks as vertical bars. (b) Schematic diagram for the crystal structure of $\mathrm{Pr_{2}Rh_{2}Ga}$.}
         \label{XRD}
\end{figure}

Fig.~1a depicts PXRD pattern measured at room temperature. For investigating the crystal structure and phase purity, the data was processes with a Rietveld refinement method using the FULLPROF software \cite{rietveld, fullprof}. The PXRD pattern of $\mathrm{Pr_2Rh_2Ga}$ along with the refinement fitting is shown in Fig.~\ref{XRD}a. The refinement results revealed that this compound crystallizes in the $\mathrm{La_2Ni_3}$-type of orthorhombic structure belonging to the $Cmca$ space group. In the structure of $\mathrm{La_2Ni_3}$, the rare-earth atom Pr occupies La site, whereas Rh and Ga occupy the two sites of Ni \cite{La2Ni3, Ce2Rh2Ga}. The obtained refinement parameters are listed in Table\ref{TABLE1}. 
A schematic diagram for the crystal structure was generated from the refinement data  by using VESTA software and is shown in Fig.~\ref{XRD}b \cite{Vesta}. The details arrangement of atoms in crystal structure is described for the similar compound of $\mathrm{Ce_{2}Rh_{2}Ga}$ \cite{Ce2Rh2Ga}.

The obtained Pr-Pr distances are ranging from 3.431 to 3.710 ~\AA{}, which is approximately twice the metallic radius of Pr element ($\mathrm{r_{Pr}}$ = 1.810~\AA{})~\cite{atomicradius}. The short interatomic distances of Pr-Rh is 2.978~\AA{}, Pr-Ga is 3.334~\AA{} and Rh-Ga is 2.547~\AA{}, which is significantly smaller than the sum of the metallic radius ($\mathrm{r_{Rh}}$ = 1.345~\AA{}) and Ga ($\mathrm{r_{Ga}}$ = 1.411~\AA)\cite{atomicradius} of two atoms. These results suggest that there is a strong bonding between these elements.

\begin{table}
\caption{\label{TABLE1} The lattice parameters and unit cell volume $\mathrm{Pr_2Rh_2Ga}$ compound obtained from the Rietveld refinements of XRD patterns for Orthorhombic phase along with the atomic coordinate positions.}
\begin{ruledtabular}
\begin{tabular}{ccccc}
\hspace{-1.5 in} a 	& \hspace{-0.0 in}  & \hspace{-0.5 in} & \hspace{-0.1 in}  & \hspace{-0.7 in}5.869(3) \AA    \\
\hspace{-1.5 in} b 	& \hspace{-0.0 in}  & \hspace{-0.5 in} & \hspace{-0.1 in}  & \hspace{-0.7 in} 9.607(2) \AA \\
\hspace{-1.5 in} c  & \hspace{-0.0 in}  & \hspace{-0.5 in} & \hspace{-0.1 in}  & \hspace{-0.7 in} 7.464(2) \AA \\
\hspace{-1.5 in} V  & \hspace{-0.0 in}  & \hspace{-0.5 in} & \hspace{-0.1 in}  & \hspace{-0.7 in} 420.80(3)  \AA$^3$ \\
\hline
\\
\hspace{0.1 in} Atomic coordinates  for $\mathrm{Pr_2Rh_2Ga}$ 	& \hspace{-0.2 in}  & \hspace{-0.2 in} & \hspace{-0.2 in}     & \hspace{-0.2 in} \\
\hline
\\
 \hspace{-1.5 in}Atom & \hspace{-2.5 in}Wyckoff & \hspace{-1.5 in}$x$ & \hspace{-0.7 in}$y$ & \hspace{-0.2 in}$z$ \\
 \hline
 & & & & \\
 \hspace{-1.5 in}Pr & \hspace{-2.5 in}$8f$ & \hspace{-1.5 in}0 & \hspace{-0.7 in}0.3385(2) & \hspace{-0.2 in}0.0981(2)\\
 \hspace{-1.5 in}Rh & \hspace{-2.5 in}$8e$ & \hspace{-1.5 in}1/4 & \hspace{-0.7 in}0.0961(1) & \hspace{-0.2 in}1/4\\
 \hspace{-1.5 in}Ga & \hspace{-2.5 in}$4a$ & \hspace{-1.5 in}0.0 & \hspace{-0.7 in}0.0 & \hspace{-0.2 in}0.0\\
\end{tabular}
\end{ruledtabular}
\end{table}

\subsubsection{Magnetic properties}
\begin{figure}
	\centering
	\includegraphics[width =3.4 in, height =2.8 in]{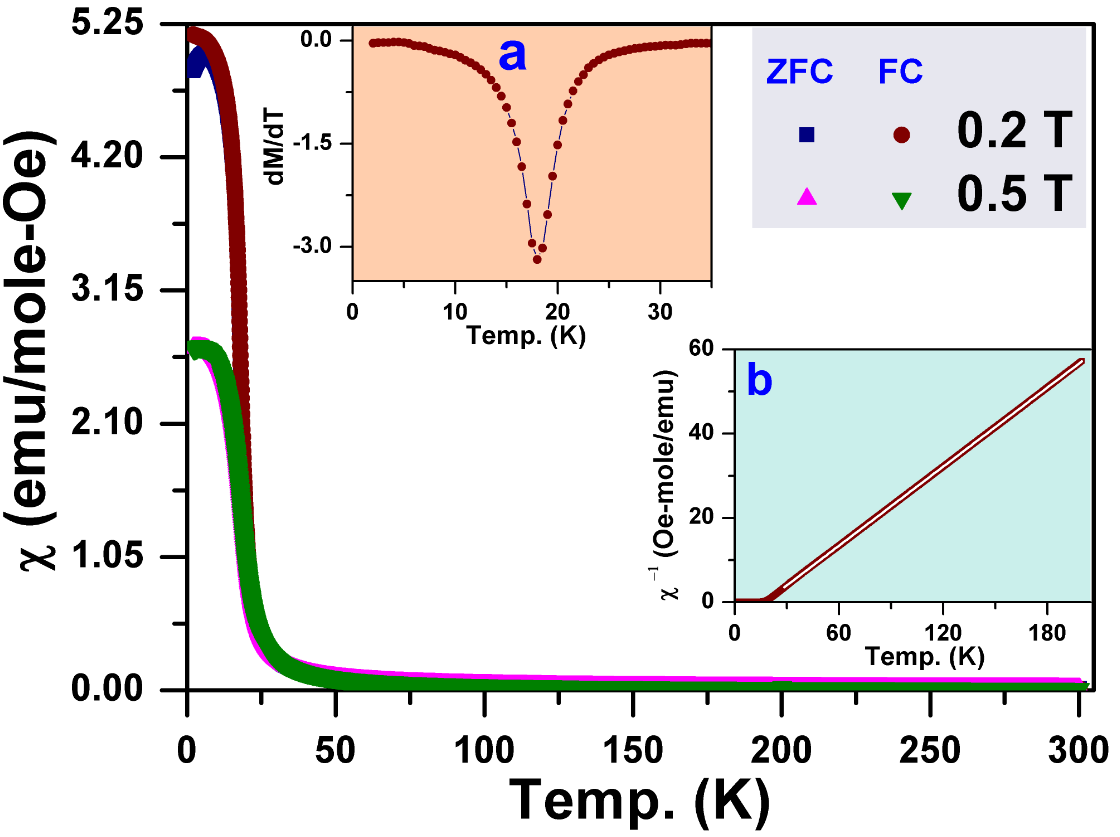}
	\caption{Temperature dependence of the dc-magnetic susceptibility $\chi(T)$ of $\mathrm{Pr_{2}Rh_{2}Ga}$ in the field-cooled (FC) and zero-field-cooled (ZFC) process. Inset (a) shows the d$M(T)$/d$T$ of FC curve under 0.2 T to estimate the transition temperature. Inset (b) temperature variation inverse magnetic susceptibility of FC curve under 0.2 T.}
	\label{MT}
\end{figure}

The temperature variation of dc$\textendash$magnetization ($M(T)$) was carried out in both zero field cooled (ZFC) and field cooled (FC) protocol in the applied external applied magnetic fields of 0.2 T and 0.5 T. In the ZFC process, the sample was cooled down to 2 K in zero field. Thereafter, a field was applied and data were recorded while warming the sample to high temperature. In FC process, the sample was cooled down in presence of magnetic field to 2 K, thereafter data was recorded upon warming from 2 K.  Fig.~\ref{MT} shows the temperature dependence of dc$\textendash$magnetic susceptibility $\chi(T)$ = $M(T)/H$,  for $\mathrm{Pr_{2}Rh_{2}Ga}$. $\chi(T)$ shows that the sample exhibits ferromagnetic behavior. The Curie temperature ${T_{C}}$ was estimated from the peak of the d$M(T)$/d$T$ of the FC magnetization curve for 0.5 T, and was found to be ${T_{C}}$ = 18 K, which is shown in inset (a) of Fig.~\ref{MT}. The ferro-
magnetic transition temperature deduced for $\mathrm{Pr_{2}Rh_{2}Ga}$
is higher than the reported for $\mathrm{Ce_{2}Rh_{2}Ga}$ \cite{Ce2Rh2Ga}.
It is also noteworthy that the ZFC and FC curves start to diverge below ${T_{C}}$, which is known as irreversibility behavior. The difference between ZFC and FC magnetization becomes negligible for the applied field of 0.5 T. This irreversibility behavior at small value of applied magnetic field probably arises due to the presence of uniaxial anisotropy associated with 4f-electron cations, short-range magnetic ordering or the domain wall pinning effect \cite{book}. However, in this case, the ZFC magnetization decreases with decrease in temperature far below ${T_{C}}$, which causes the irreversibility behavior. This point towards domain wall pinning due to the magnetocrystalline anisotropy of the $\mathrm{Pr^{3+}}$ ions, as a probable cause of the irreversibility behavior in the compound \cite{Pr2Rh3Ge}.

The temperature variation of inverse dc-magnetic susceptibility, $\mathrm{\chi^{-1}}(T)$ for H = 0.2 T is depicted in inset (b) of Fig.~\ref{MT}. The $\mathrm{\chi^{-1}}(T)$ graph shows a linear behavior at temperatures above 20 K, which follows the Curie$\textendash$Weiss law; $\chi(T) = \mathrm{C}/(\textit{T}- \theta_{p}$), where C is the Curie constant and is defined as $\mathrm{C = N \mu^{2}_{eff}/3k_{B}}$. $\mathrm{\theta_{p}}$ is the paramagnetic Weiss temperature. A linear fit to the inverse susceptibility in the data yields $\mathrm \theta_{p}$ = 17 K. The positive value of $\mathrm{\theta_{p}}$ indicates the presence of a strong ferromagnetic exchange interaction in the system. The calculated value of the effective moment ($\mathrm{\mu_{eff}}$) is 3.57 $\mathrm{\mu_{B}}$/$\mathrm{Pr^{3+}}$, which is very close to the theoretical value for a free trivalent $\mathrm{Pr^{3+}}$, $\mathrm{g_{J}[J(J + 1)]^{1/2}}$ = 3.58 $\mathrm{\mu_{B}}$ for J = 4. This result indicates that 4f shell electrons of $\mathrm{Pr^{3+}}$ ions are the predominant magnetic species.

\begin{figure}
	\centering
	\includegraphics[width =3.4 in, height =2.8 in]{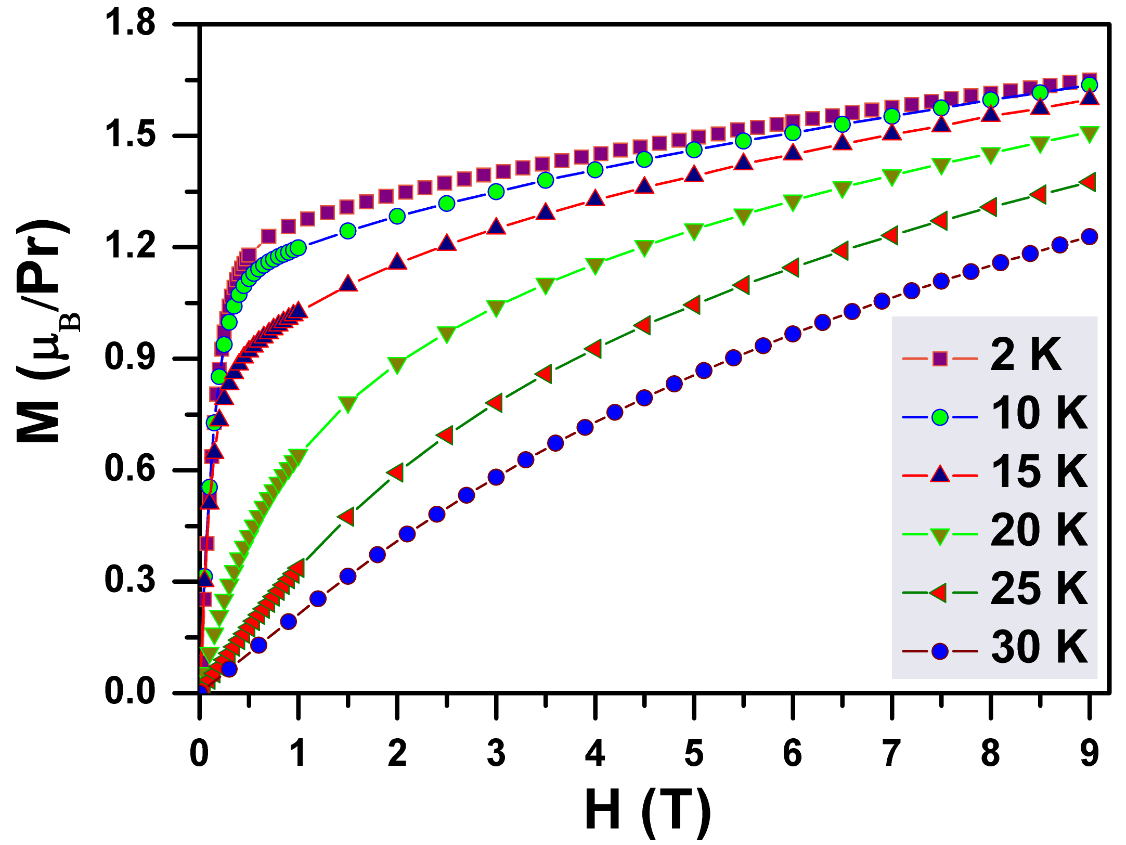}
	\caption{Isothermal magnetization at different temperatures.}
	\label{MH}
\end{figure}

Fig.~\ref{MH} shows the field-dependent isothermal magnetization of $\mathrm{Pr_{2}Rh_{2}Ga}$ compound. A set of isothermal magnetization $M(H)$ curves were measured with magnetic field as high as 9 T. The measurements were carried out at different temperatures from 2 to 30 K ( for both above and below $T_C$). It is also observed that the magnetization does not reach saturated value even at high field of $H$ = 9 T at low temperature of 2 K. The spontaneous magnetization of the compound at 2 K is 1.28 $\mathrm \mu_{B}$/Pr. The obtained magnetization of $\mathrm{Pr_{2}Rh_{2}Ga}$ is smaller than the theoretical value of $\mathrm{Pr^{3+}}$ free ions, gJ = 3.2 $\mathrm \mu_{B}$/Pr. The magnetic moment per Pr ions at 9 T for 2 K, is found to be 1.65 $\mathrm \mu_{B}$/Pr, which is also two times less than the saturated moment value expected for parallel alignment of free $\mathrm{Pr^{3+}}$. The low saturation value may be attributed to the influence of anisotropy due to the CEF surrounding Pr ions \cite{Pr2Rh3Ge}.

\subsubsection{Heat capacity}
\begin{figure}
      \centering
        %\begin{tabular}{c}
        %example1\\
        \includegraphics[width =3.5 in, height =2.8 in]{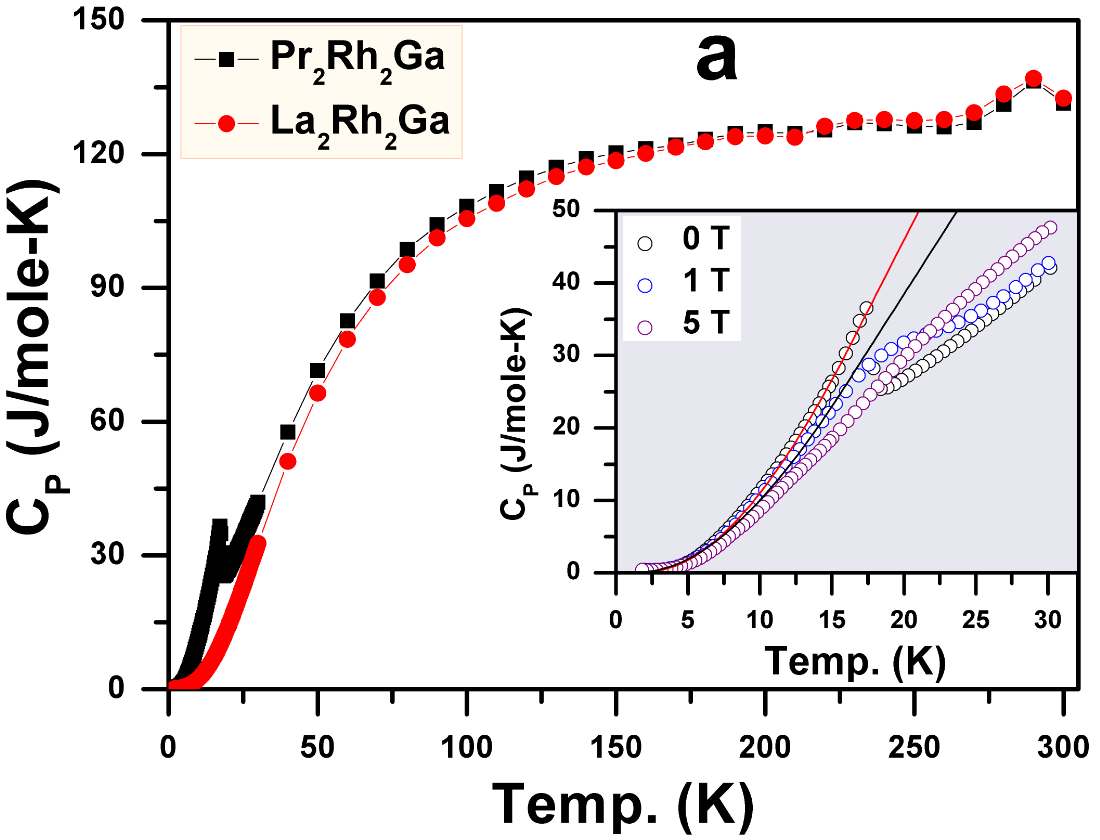}
       %\end{tabular}

      \vspace{1.0em}
        \begin{tabular}{cc}
        %example2  & example3 \\
        \includegraphics[width =1.70 in, height =1.5 in]{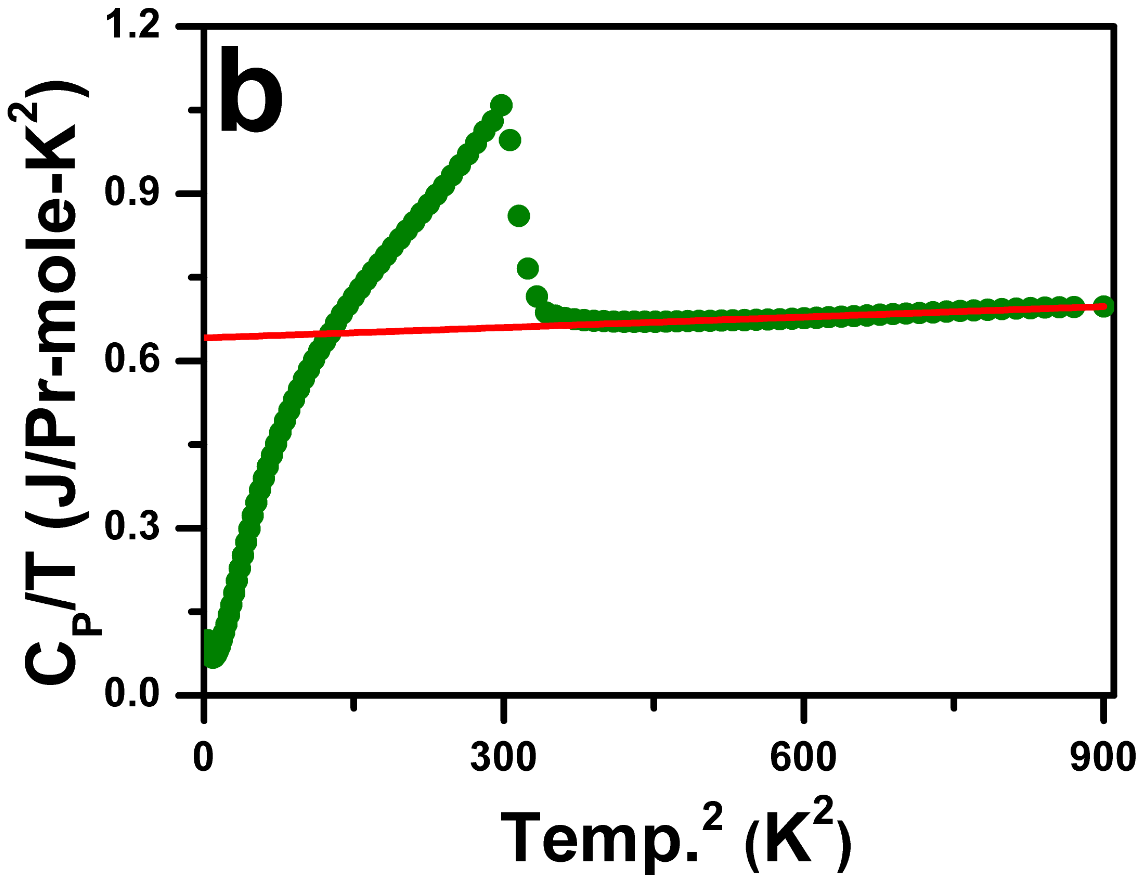}
         &
         \includegraphics[width =1.70 in, height =1.5 in]{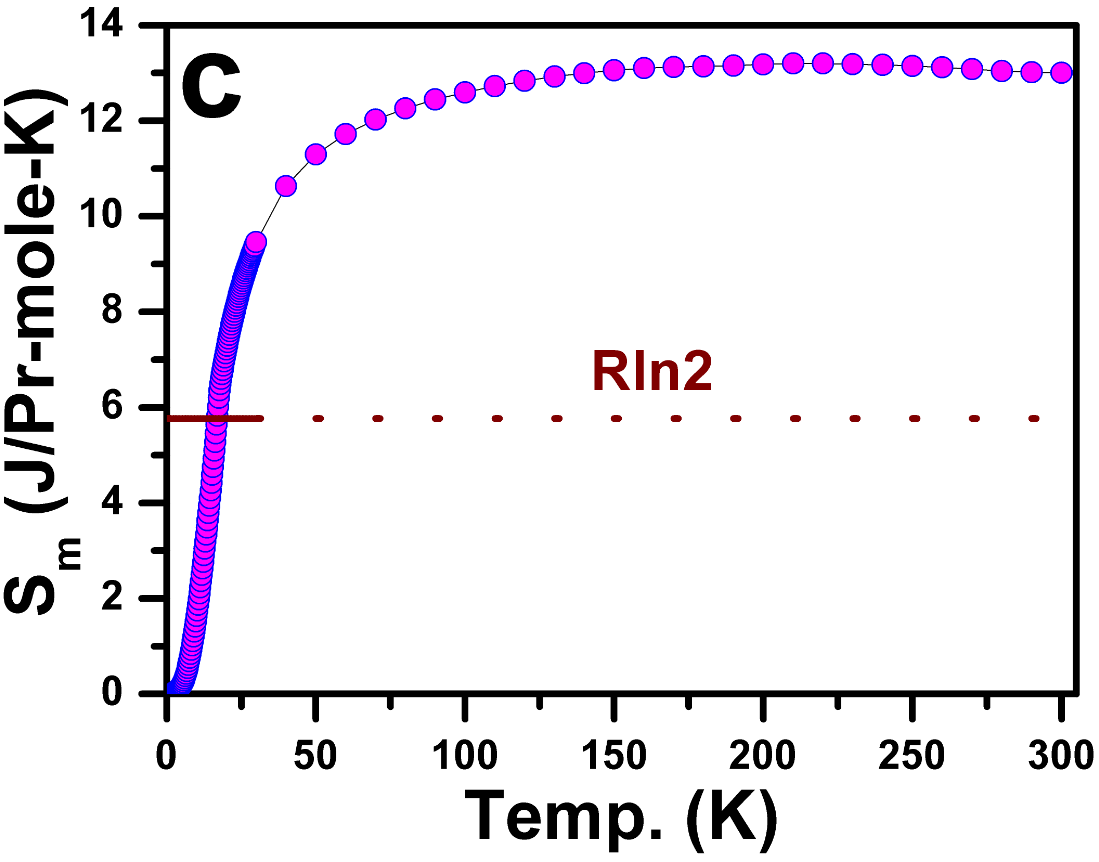}
         \end{tabular}
         \caption{(a) Temperature dependence of the heat capacity $C_P(T)$ of $\mathrm{Pr_{2}Rh_{2}Ga}$ measured in zero field. $C_P(T)$ data of the non-magnetic reference compound $\mathrm{La_{2}Rh_{2}Ga}$ \cite{Ce2Rh2Ga}. Inset: temperature dependent heat capacity under different magnetic field. (b) The low-T part of $C_P(T)$/$T$ as a function of $T^2$ together with fitting for evaluating the Sommerfeld coefficient and Debye temperature. (c) The calculated 4f-electron entropy as a function of temperature.}
         \label{CP}
\end{figure}

Fig.~\ref{CP}a depicts the $\mathrm{\textit{C}_{p}}(\textit{T})$ for $\mathrm{Pr_{2}Rh_{2}Ga}$ in solid black symbols and non-magnetic reference compound,  $\mathrm{La_{2}Rh_{2}Ga}$ in solid red symbols. At room temperature, $\mathrm{\textit{C}_{p}}(\textit{T})$ is found to reach a value of 128 J/$\mathrm{(mole.K)}$. These values deviate only 3\% from the Dulong-Petit value under the formula $\mathrm{\textit{C}_{p}}$ = 3nR = 124.7 J/$\mathrm{(mole.K)}$, where n is the number of atoms per formula unit (in this case n = 5) and R is the universal gas constant. At low temperatures, $\mathrm{\textit{C}_{p}}(\textit{T})$ shows a $\lambda$-type anomaly, which marks the ferromagnetic ordering in the compound. The discontinuous jump with $\lambda$-shape anomaly of heat capacity result indicates that the compound undergoes second order ferromagnetic phase transition \cite{R2NiSi3}.

At low temperature, the heat capacity of the metal is contributed by both electronic and phononic contribution, i.e $\mathrm{\textit{C}_{p}}$ = $\mathrm{\textit{C}_{el}}$ + $\mathrm{\textit{C}_{ph}}$. The free electron nature of the quasi-particle interactions in an electronic system is indicated by the Sommerfeld coefficient($\gamma$) \cite{CPbook}. The value of $\gamma$ can be obtained from the linear fit of the $\mathrm{\textit{C}_{p}}/ \textit{T}$ $vs.$ ${T^{2}}$ plot by assuming the general expression $\mathrm{\textit{C}_{p}}=\gamma\textit{T}+\beta{\textit{T}^{3}}$. In order to perform the fit, the lowest available paramagnetic temperature region was used and is shown in Fig.~\ref{CP}b along with the fitted line (red line). The best fit data yields $\gamma$ = 640 mJ/$\mathrm{(Pr.mole.K^2)}$. The obtained $\gamma$ value is compared with other heavy fermions Pr-based ternary compounds $viz.,$ 286 mJ/$\mathrm{(Pr.mole.K^2)}$ for $\mathrm{PrRhSn_3}$ \cite{PrRhSn3}, 315 mJ/$\mathrm{(Pr.mole.K^2)}$ for $\mathrm{Pr_2Rh_3Ge}$ \cite{Pr2Rh3Ge}, 716 mJ/$\mathrm{(Pr.mole.K^2)}$ for $\mathrm{Pr_3Rh_4Sn_{13}}$ \cite{Pr3Rh4Sn13}, 300 mJ/$\mathrm{(Pr.mole.K^2)}$ $\mathrm{Pr_3Ru_4Ge_{13}}$ \cite{Pr3Ru4Ge13} and 300 mJ/$\mathrm{(Pr.mole.K^2)}$ for $\mathrm{PrV_2Al_{20}}$ \cite{PrV2Al20}. The value obtained here is comparable to the reported value for the Pr-based ternary compounds. The Debye-temperature($\mathrm \theta_{D}$) was also extracted using the fitted value of $\beta$ and yield $\mathrm \theta_{D}$ = 313 K.

The inset to Fig.~\ref{CP}a present the low-temperature $\mathrm{\textit{C}_{p}}(\textit{T})$ data in the range between 2 K and 30 K, which were  measured under the different values of magnetic field $viz.,$ 0, 1 and 5 T. One can see that the peak associated with the phase transition shifts towards higher temperatures with applied magnetic field and also suppresses at high magnetic field of 5 T. This feature is commonly seen in ferromagnetic compounds \cite{CPbook}. Below $T_C$, the $C_{p}(\textit{T})$ data can be described with the following formula;

\begin{eqnarray}
 C_{\mathrm p}(T)=\gamma_e T + B T^{3/2} \exp^{(\frac{-\Delta}{T})},
\end{eqnarray}
where $\gamma_e$ is the electronic contribution to the heat capacity
in the ordered state. The second term $B T^{3/2} exp^{(\frac{-\Delta}{T})}$
represents the spinwave contribution for a ferromagnet with an energy gap $\Delta = E_g/K_B$ in the magnon spectrum of the heat capacity. The best fit on the experimental data is shown in inset of Fig.~\ref{CP}a with solid lines and yielded parameters: $\gamma_e$ = 0.029(3) J/(mole-$K^2$), B = 0.794(3) J/(mole-$K^2$), $\Delta$ = 8.47(3) K in zero field, and $\gamma_e$ = 0.053(4) J/(mole-$K^2$), B = 0.616(3) J/(mole-$K^2$), $\Delta$ = 7.16(2) K in 1 T. The values of $\Delta$ are smaller than $T_C$, which is in good agreement with the previously reported $\mathrm{RE_{2}T_{2}X}$ system \cite{{Nd2Pt2In,Pr2Pt2In}}. 
 
The magnetic entropy $\mathrm{\textit{S}_{m}}$ has been estimated by integrating ($\mathrm{\textit{C}_{m}}(\textit{T})/ \textit{T}$) as a function of $\textit{T}$ and is shown in Fig.~\ref{CP}c. Here, $\mathrm{\textit{C}_{m}}(\textit{T})$ was estimated by subtracting of $\mathrm{\textit{C}_{p}}$ of $\mathrm{La_2Rh_2Ga}$ from $\mathrm{\textit{C}_{p}}$ of $\mathrm{Pr_2Rh_2Ga}$. As seen from Fig.~\ref{CP}c, the $\mathrm{\textit{S}_{m}}$ released at ${T_{C}}$ is very close to the value of Rln2. This reflects that the Pr-has doublet magnetic ground state. It is also noticed that the $\mathrm{\textit{S}_{m}}$ gradually increases with increasing temperature and fully saturate at 100 K, which is much higher than the phase transition temperature. The obtained saturation $\mathrm{\textit{S}_{m}}$ value at > 100 K is only about 70 $\%$ of the full Rln(2J+1), J = 4 entropy.

\subsubsection{Electrical resistivity}

\begin{figure}
	\centering
	\includegraphics[width =3.45 in, height =3.0 in]{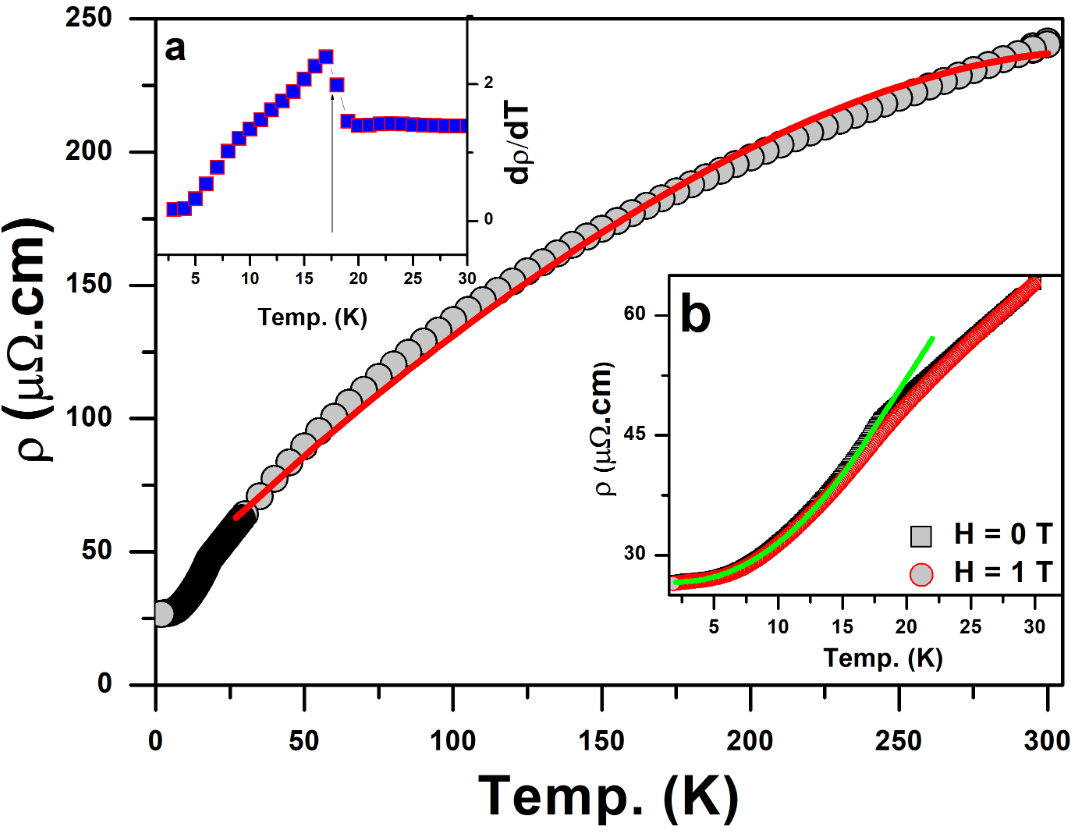}
	\caption{Temperature dependence of resistivity under zero magnetic field and the red solid line represents the fit of Eq.~(2) to the experimental data.. The upper inset~(a) d$\rho$/dT as a function of temperature. The arrow indicates the critical temperature $T_C$ associated with $\rho$. Inset~(b) solid symbols are low-temperature $\rho(T)$ measured under field of 0 and 1 T and the solid line represents the fit of Eq.~(3) to the experimental data.}
	\label{resistivity}
\end{figure}

The $\rho(T)$ of $\mathrm{Pr_{2}Rh_{2}Ga}$ is presented in Fig.~\ref{resistivity}. It is observed from Fig.~\ref{resistivity} that the resistivity gradually decreases with decreasing temperature and shows well defined kink at the magnetic transition. The compound shows overall metallic behavior. It is also noticed that the resistivity data shows a strong curvature in the paramagnetic region (above 50 K), which is associated with the substantial electron-phonon interaction strength at high temperature and also the scattering effects of the conduction electrons on disordered magnetic moments in combination with the CEF effect \cite{Pr2Rh3Ge}. At low temperature, s-d interband scattering of the conduction electrons is related to a Mott term. In the paramagnetic region $T~ \textgreater$ 50 K, experimental data of $\rho(\textit{T})$ was derived using the following the Bloch - Gr\"{u}neisen - Mott formula \cite{R-Debey1, R-Debey2}.
\begin{eqnarray}
\nonumber
\rho(T)=\rho_{0}+\frac{4A_{0}}{\theta_{R}}\left(\frac{T}{\theta_{R}}\right)^{5}\int\limits_{0}^{\theta_{R}/T}\frac{x^{5}dx}{(e^{x}-1)(1-e^{x})}\\
-KT^{3},
\end{eqnarray}
where $\rho_{0}$ is the residual resistivity typically associated with metallurgical defects. ${A_{0}}$ is the electron-phonon coupling constant. $\theta_{R}$ is the Debye temperature,  which represents the characteristic energy scale of lattice vibrations. In the last term, K of Eq.(2) is known as the Mott coefficient, which is a characteristic of s$\textendash$d interband scattering. The best fits of Eq.~(2) with solid line to the experimental $\rho(\textit{T})$ vs $T$ data  yielded:   $\rho_{0}$ = 51.29(2) $\mathrm \mu.\Omega.cm$; $\mathrm \theta_{R} = 102.84(2)~K$; $\mathrm{\textit{A}_{0}}= 9.25(4)\mu.\Omega.cm.K^{-3}$ and   $\textit{K} = 2.78(3)\times 10^{-6} \mathrm \mu.\Omega.cm.K^{-3}$. 

The inset (b) of Fig.~\ref{resistivity} depicts an expanded view of $\rho(\textit{T})$ data at low temperature. The resistivity decreases sharply below the phase transition which is due to spin disorder scattering of the conduction electrons have been quenched by the magnetic ordering of the spins. The magnetic transition temperature was also evaluated from $\rho(\textit{T})$ result by taking the derivative of $\rho$ with respect to $T$. The expanded region of d$\rho$/d$T$ curve at low temperature is presented in inset (a) of Fig.~\ref{resistivity}. According to Sato criterion \cite{Sato}, the value of ${T_{C}}$ = 18 K was estimated from the midpoint of the anomaly in the d$\rho$/d$T$ curve, and is marked with an arrow inside of the figure. The magnetic phase transition temperature in $\rho$ is consistent with the results of $\chi(T)$ and $\mathrm{\textit{C}_{p}}(\textit{T})$ data.

In order to analyse the characteristic features for resistivity associated with the magnetic ordered state, the temperature dependence $\rho(\textit{T})$ was measured at low temperature under magnetic field value of 1 T and is plotted in the inset (b) of Fig.~\ref{resistivity}. As seen from the inset (b) of Fig.~\ref{resistivity}, the anomaly at the transition temperature is suppressed with the application of magnetic field, which is generally seen in a ferromagnetically ordered system. The scattering of the conduction electrons in terms of magnons can be explained though the spin-wave excitation. Below ${T_{C}}$, the temperature variation of $\rho$ under zero-magnetic field data was described by using spin-wave excitation of Eq.~(3) \cite{R-spinwave}. 
 
\begin{eqnarray}
\rho(T) = \rho_{FM}+A\Delta_{R}T\left[1+2\frac{T}{\Delta_{R}}\right]\exp\left[-\frac{\Delta_{R}}{T}\right],
\end{eqnarray}
where $\rho_{FM}$ represents the residual resistivity of the magnetic ordered state, A is a material constant which depends on the spinwave stiffness, and $\Delta_{R}$ is associated with the energy gap in the ferromagnetic magnon spectrum. A best fit of Eq.(3) on the experimental result with solid line yields fitting parameters: $\rho_{FM}$ = 26.6(3) $\mathrm{\mu\Omega.cm}$ , A = 0.015(2) $\mathrm{\mu\Omega.cm.K^{-2}}$ and $\Delta_{R}$ = 9.01(3) K.

\begin{figure}
	\centering
	\includegraphics[width =3.5 in, height =2.8 in]{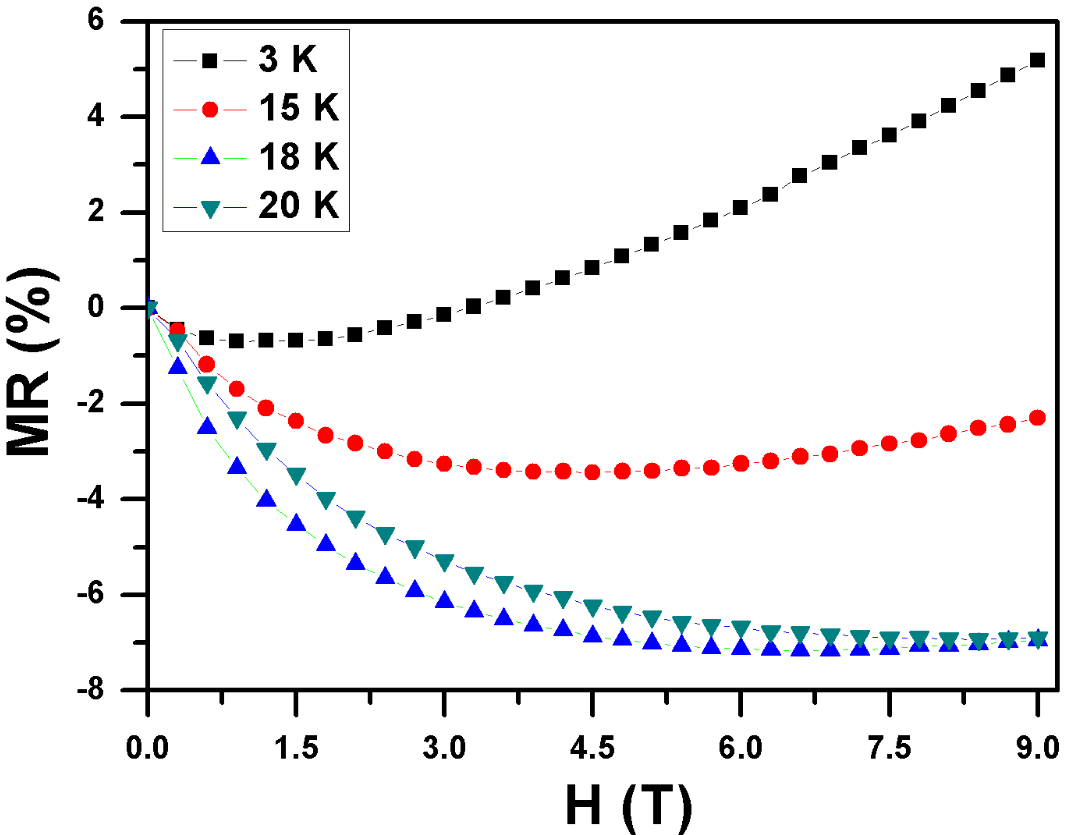}
	\caption{Field dependence of the magnetoresistance isotherms of
	 $\mathrm{Pr_2Rh_2Ga}$ at different temperature near $T_C$.}
	\label{MR}
\end{figure} 

In order to investigate the magnetoresistance (MR) behavior of $\mathrm{Pr_2Rh_2Ga}$, isothermal magnetic field dependence of resistivity was measured at different temperatures for both below and above $T_C$. MR was calculated from the isothermal magnetic field dependence of resistivity curve using the following formula of Eq.(4). The obatined MR variation with magnetic field was plotted in Fig.~\ref{MR}. 

\begin{eqnarray}
MR = \frac{\rho(H,T)-\rho(0,T)}{\rho(0,T)}~100 \%
\end{eqnarray}

It is seen from Fig.~\ref{MR} that isothermal MR shows both positive and negative values at high fields of 9 T depending on the temperature. As seen the data for lowest temperature of 2 K, the magnitude of MR is positive above $H$ = 3 T and gradually increases with increasing field. However, negative MR is seen for $T$ = 15 K and higher. This result indicates that both positive and negative MR can exists below the transition temperature. The negative MR in ferromagnetic region can be attributed to reduction in spin-disorder resistivity. However, the positive MR in ferromagnetic compound can be understood in-terms of the Lorentz’s force, which causes the classical modification of the electron trajectory \cite{Pr2Rh3Ge}. The MR at 20 K just above the $T_C$ shows negative behavior, which may suggest the presence of weak ferromagnetic correlations at 20 K. Similar behavior of MR properties below and above ${T_{C}}$ was also reported in the case of $\mathrm{Pr_{2}Rh_{3}Ge}$ and $\mathrm{CeIr_{2}B_{2}}$ compounds \cite{Pr2Rh3Ge, CeIr2B2}.

\subsection{\label{sec:level2}Magnetocaloric effect}

\begin{figure}
	\centering
	\includegraphics[width =3.4 in, height =3.0 in ]{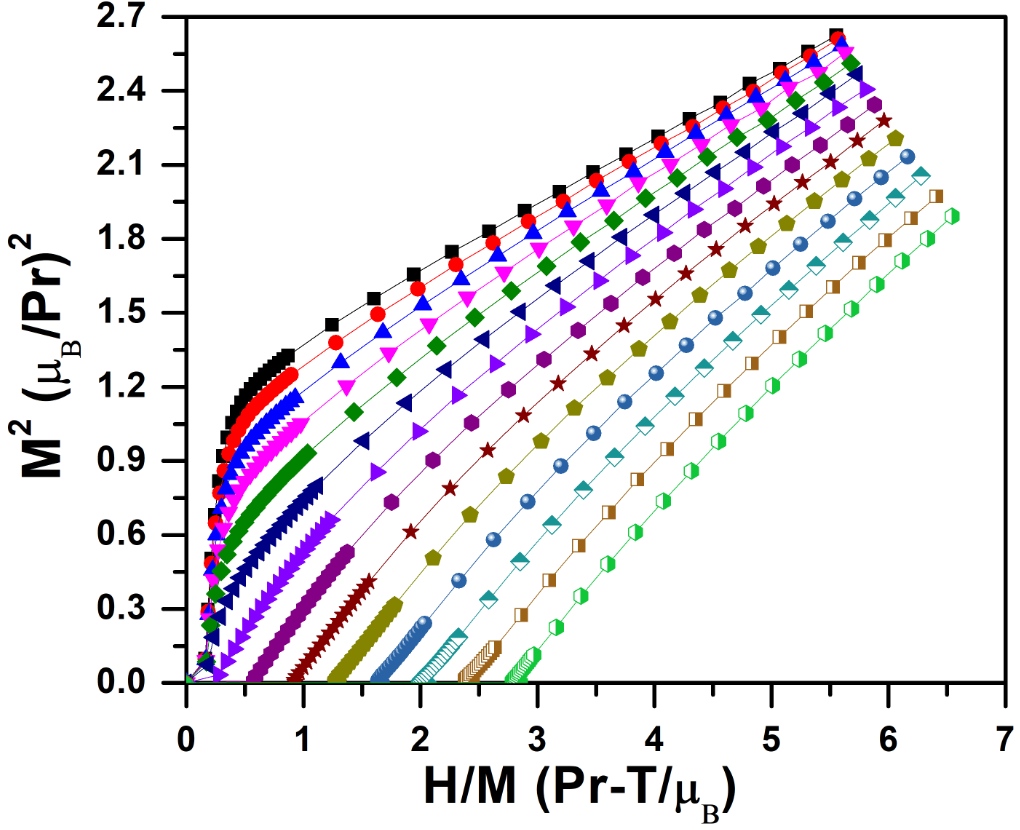}
	\caption{Arrott plots ($M^{2}~vs.~H/M$) derived from the magnetization isotherms in the temperature range of 12 - 25 K with a step of 1 K.}
	\label{AP}
\end{figure}

The magnetocaloric effect of the $\mathrm{Pr_{2}Rh_{2}Ga}$ is explored through isothermal magnetization and heat capacity measurements. The order of the phase transition is also confirmed from Arrott plots. Arrott plots ($M^{2}~vs.~H/M$) were performed from the $M(H)$ curves. Fig.~\ref{AP} shows the isotherms of $M^{2}~vs.~H/M$ plot in the temperature range of 12 K to 25 K (near the transition temperature region). According to Banerjee criterion, a positive slope in  the $M^{2}~vs.~H/M$ plot implies that the system possesses a second order magnetic transition \cite{Banerjee}.

Arrott plots are commonly used to determine the nature of a magnetic phase transition based on the isothermal magnetization data. However, determining the magnetic phase transition from Arrott plot has also some limitation for the shortcomings such as meta-magnetic transition, demagnetization field and domain wall pinning effect \cite{Ramesh2}. Therefore, Bonilla and Franco \cite{PRB-Franco} have suggested one more method to determine the second order of phase transition by employing $\rm -\Delta \textit{S}_{M}~vs.~\textit{T}$ curve, which will be discussed for this  $\rm{Pr_{2}Rh_{2}Ga}$ compound.

The $\rm \Delta \textit{S}_{M}$ was calculated from isothermal magnetization curve using the following Maxwell relation \cite{book-Tishin}: 

\begin{eqnarray}
\rm \Delta \textit{S}_{M} (\textit{T,H}) = \int\limits_{0}^{\textit{H}} \left(\frac{\partial \textit{M}}{\partial \textit{T}}\right)d\textit{H}.
\end{eqnarray}

\begin{figure}
	\centering
	\includegraphics[width =3.4 in, height =3.1 in ]{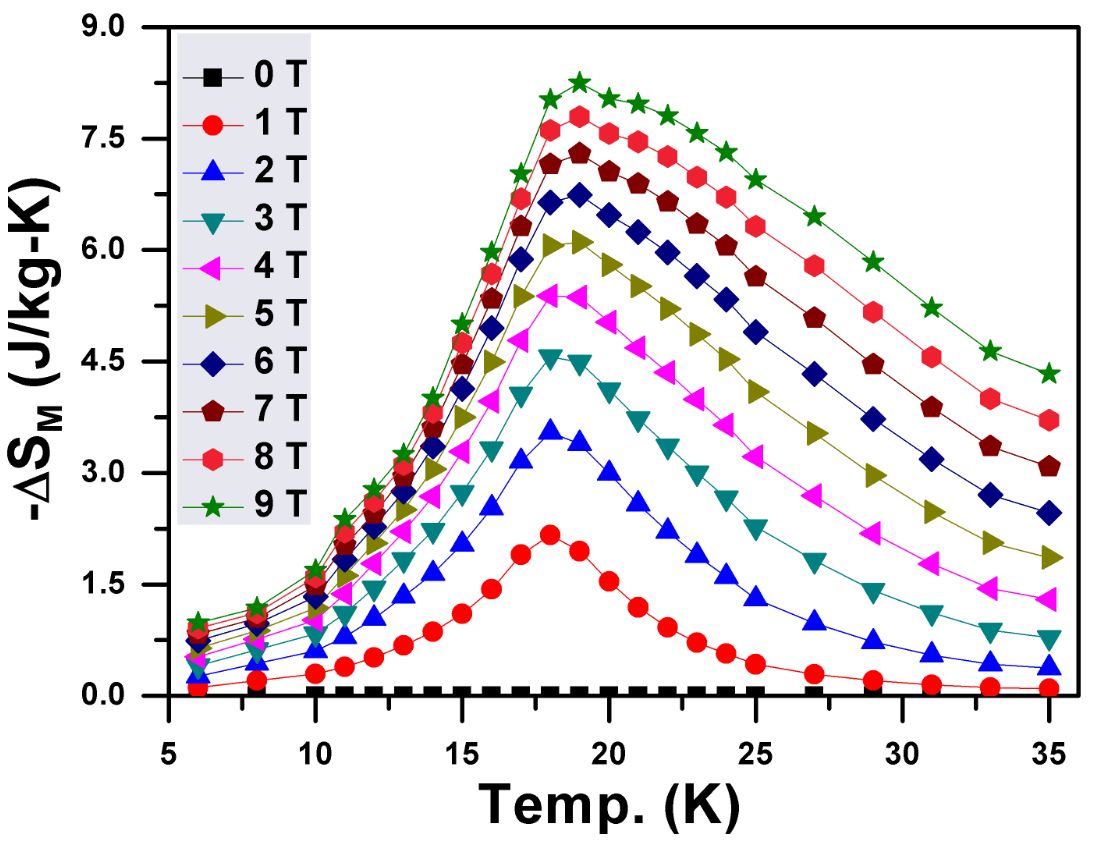}
	\caption{Temperature variation magnetic entropy changes ($\rm \Delta \textit{S}_{M}$) for different values of changing magnetic field}
	\label{MCE}
\end{figure}

Fig.~\ref{MCE} shows the variation of $\rm -\Delta \textit{S}_{M}$ as a function of temperature for different values of magnetic field. It is seen that there is no tendency of saturation in $\rm -\Delta \textit{S}_{M}$ values even at applied magnetic field strength of $9 \rm ~T$. It is found that the maximum values of $\rm -\Delta \textit{S}_{M}$ are  $6.1,~ 7.3$ and $8.2~ \rm J/kg. K$ for the change of magnetic field of $0\textendash5~\rm T$, $0\textendash7~\rm T$, and the $0\textendash9~\rm T$, respectively. The obtained $\rm -\Delta \textit{S}_{M}$ values are comparable with the reported value of other Pr based ternary compounds $viz.$, $\rm -\Delta \textit{S}_{M}$ = $5.8~ \rm J/kg. K$ for 7 T of $\rm{Pr_{2}Pt_{2}In}$ and $\rm -\Delta \textit{S}_{M}$ = $6.1~ \rm J/kg. K$ for 5 T of $\rm{Pr_{6}Co_{2}Si_{3}}$ \cite{Pr2Pt2In, Pr6Co2Si3}. The observed values are also compared with the some other reported $\mathrm{RE_2T_2X}$ compound in Table.~\ref{compare}. As seen from the Table.~\ref{compare}, the obtained value of $\mathrm{Pr_2Rh_2Ga}$ is also quite large and comparable to those of the reported refrigerant materials around corresponding transition temperature. From this comparison, one can say that the present  $\rm{Pr_{2}Rh_{2}Ga}$ compound is in a materials class that may profitably be exploited from MCE.

\begin{table}[ht]
\caption{The transition temperature, the maximum values of
magnetic entropy change ($\rm \Delta \textit{S}_{M}$), and refrigeration capacity (RC) under the field change of $0\textendash5~\rm T$ for some rare-earth compounds of $\mathrm{RE_2T_2X}$.}
\centering 
\begin{tabular}{ccccc}
\hline\hline
\\  
Method&  \hspace{0.2cm}$\rm{\textit{T}_{N}}$/$\rm{\textit{T}_{C}}$& \hspace{0.7cm}$\rm -\Delta \textit{S}_{M}$& \hspace{0.2cm}RC& \hspace{0.3cm}Ref  \\ [0.5ex]
       & (K) &\hspace{0.5cm}(J/kg.K) & \hspace{0.3cm}(J/kg)&     \\ [0.5ex]
\hline
\\ 
\hspace{0.0cm}$\rm{Nd_{2}Pt_{2}In}$& \hspace{0.3cm}16& \hspace{0.3cm}5.01& \hspace{0.3cm}---& \hspace{0.6cm}\cite{Nd2Pt2In} \\
\hspace{0.0cm}$\rm{Gd_{2}Ni_{2}Sn}$& \hspace{0.3cm}75& \hspace{0.2cm}4.6&   \hspace{0.3cm}---& \hspace{0.6cm}\cite{Gd2Ni2Sn} \\ 
\hspace{0.0cm}$\rm{Er_{2}Co_{2}Al}$& \hspace{0.3cm}32& \hspace{0.3cm}5.9&  \hspace{0.3cm}120& \hspace{0.6cm}\cite{Er2Co2Al} \\
\hspace{0.0cm}$\rm{Gd_{2}Cu_{2}Cd}$& \hspace{0.3cm}120& \hspace{0.3cm}7.8&  \hspace{0.3cm}234&  \hspace{0.6cm}\cite{Gd2Cu2Cd}  \\
\hspace{0.0cm}$\rm{Dy_{2}Co_{2}Ga}$& \hspace{0.3cm}55& \hspace{0.3cm}6.2&  \hspace{0.3cm}114&  \hspace{0.6cm}\cite{Dy2Co2Ga} \\
\hspace{0.0cm}$\rm{Pr_{2}Rh_{2}Ga}$& \hspace{0.3cm}18& \hspace{0.3cm}6.1& \hspace{0.3cm}70&  \hspace{0.6cm}This work  \\ [1ex]
\hline 
\end{tabular}
\label{compare} 
\end{table}

\begin{figure}
	\centering
	\includegraphics[width =3.45 in, height =3.3 in ]{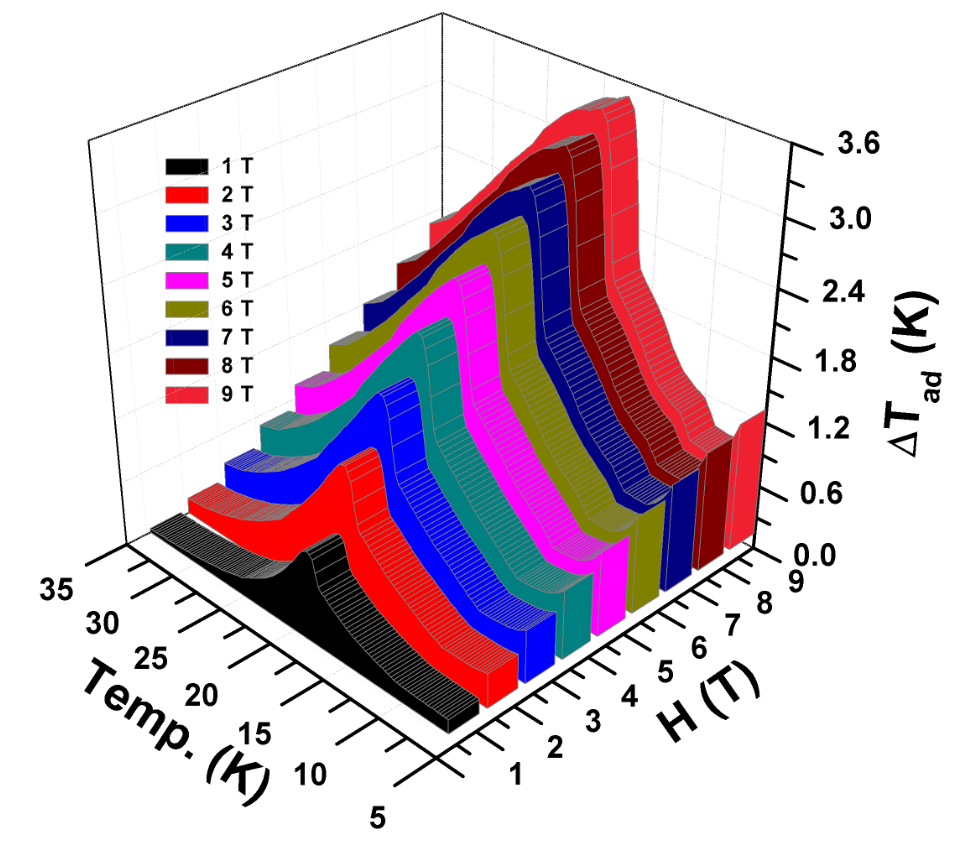}
	\caption{Temperature variation of the adiabatic temperature change ($\rm \Delta \textit{T}_{ad}$) for different values of magnetic field.}
	\label{Adiabatic}
\end{figure}

Another important parameter to determine potential of a MCE material is the $\rm \Delta \textit{T}_{ad}$. The $\rm \Delta \textit{T}_{ad}$ was calculated from temperature dependent zero-field heat capacity data (is shown in Fig.~\ref{CP}) and obtained $\rm -\Delta \textit{S}_{M}$, using Maxwell’s relation by the following formula \cite{book-Tishin}:

\begin{eqnarray}
\Delta \textit{T}_{\rm ad} \approx \frac{T}{\textit{C}_{\rm p}} |\rm \Delta \textit{S}_{M}| .
\end{eqnarray}
Fig.~\ref{Adiabatic} shows the temperature variation of $\rm \Delta \textit{T}_{ad}$ for the change of magnetic field up to 9 T. The maximum value of $\rm \Delta \textit{T}_{ad}$ was found to be 3.5 K for the change of field $0\textendash9~\rm T$. This estimated value of $\rm \Delta \textit{T}_{ad}$  is quite good as compared to some of $\rm{RE_{2}T_{2}X}$ ternary compounds for magneto-refrigerant application \cite{R2T2X}.

Additionally, the quality factor of MCE materials is the refrigeration capacity (RC) which evaluates the magnetic cooling efficiency. RC is an indirect measurement of heat transfer in an ideal MCE cycle between the cold and hot reservoirs. The RC of $\rm{Pr_{2}Rh_{2}Ga}$ was estimated from $\rm -\Delta \textit{S}_{M}~vs.~\textit{T}$ curve. As suggested by  Pecharskya and Gschneidner \cite{RC}, the RC value is estimated from the area under the curve by integration using following the formula:

\begin{eqnarray}
RC = \int\limits_{T_{1}}^{T_{2}} \left(\rm -\Delta \textit{S}_{M}\right) d\textit{T},
\end{eqnarray}
where $T_{1}$ and $T_{2}$ are the temperatures corresponding to both sides of the half-maximum value of the $\rm -\Delta \textit{S}_{M}(T)$ peak. It is found that the RC values gradually increase with increasing field. The values of RC are 70 J/kg, 100 J/kg and 135 J/kg for a change of field $0\textendash5~\rm T$, $0\textendash7~\rm T$ and $0\textendash9~\rm T$  respectively. As seen from Table.~\ref{compare}, our observed values for the present $\rm{Pr_{2}Rh_{2}Ga}$ compound is comparable even larger than those of some reported magnetic caloric materials

\begin{figure}
  	\centering
  	\includegraphics[width =3.45 in, height =3.1 in ]{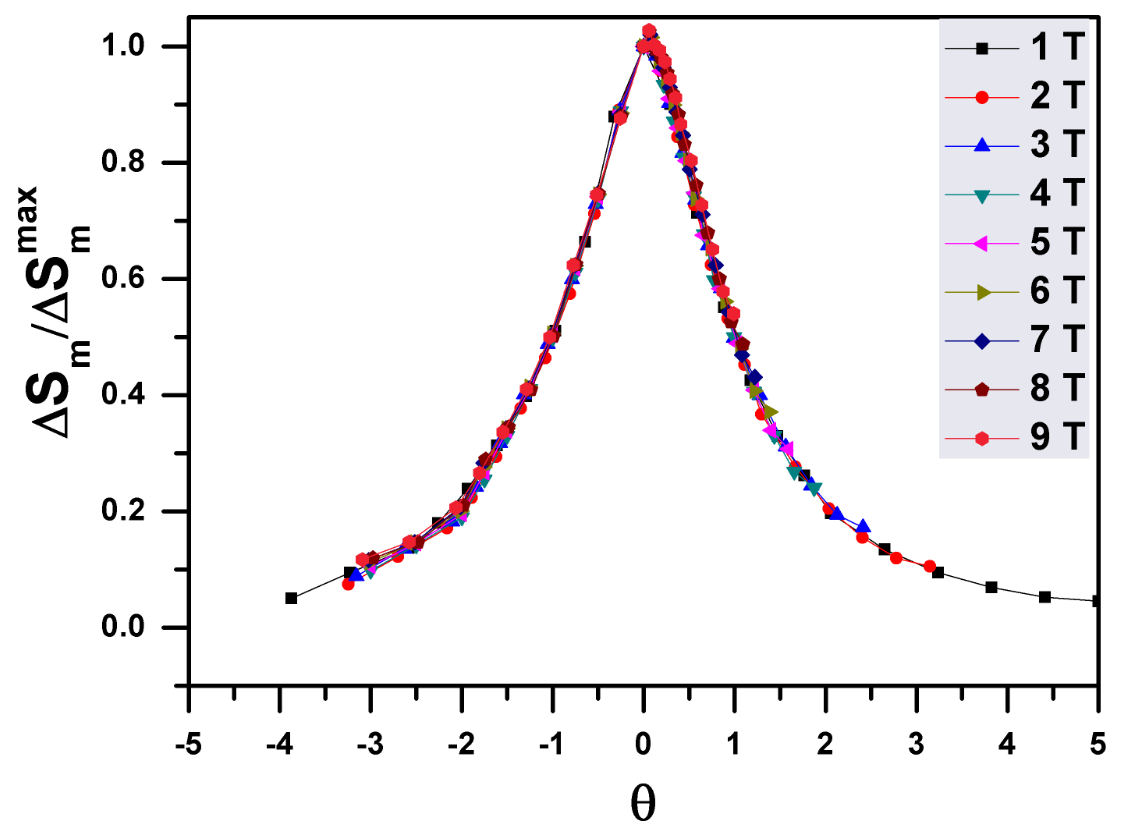}
  	\caption{Normalized entropy change ($\rm \Delta \textit{S}_{M}$/$\Delta \textit{S}^{max}_{\rm M}$) as a function of the rescaled temperature($\theta$) for the selected values of applied field}
  	\label{Scaling}
\end{figure}
  
 Magnetocaloric effect of a magnetic material also depends on the order of its magnetic phase transition. In order to get more confirmation for the second order magnetic phase transtion of $\rm{Pr_{2}Rh_{2}Ga}$, we have used universal scaling plots for the magnetic entropy changes \cite{PRB-Franco}. The universal scaling plot is derived from the $\rm -\Delta \textit{S}_{M}$ vs. $\textit{T}$ curve. The normalized entropy change $ \Delta \textit{S}_{\rm M}$/$\Delta \textit{S}^{max}_{\rm M}$  (where $\Delta \textit{S}^{max}_{\rm M}$ is the maximum entropy change) against  rescaled temperature ($\theta$) for below and above $\rm{\textit{T}_{C}}$ is plotted for different fields and is shown in Fig.~\ref{Scaling}. The rescaled temperature $\theta$ below and above $\rm{\textit{T}_{C}}$ as defined in the following equation
  
 \begin{eqnarray}
 \nonumber
\theta = -(T-T _{\rm C)})/(T_{r_1}-T_{\rm C)}\\
 (T-T_{\rm C)})/(T_{r_2}-T_{\rm C)}),
 \end{eqnarray}
where $\rm{\textit{T}_{r1}}$ and $\rm{\textit{T}_{r2}}$ is the temperature corresponding to half of the value of $\rm \Delta \textit{S}_{M}$/$\Delta \textit{S}^{max}_{\rm M}$ at $T$ \textless $\rm{\textit{T}_{C}}$ and $T$ \textgreater $\rm{\textit{T}_{C}}$, respectively. As seen from Fig.~\ref{Scaling}, the normalized entropy change with respect to rescaled temperature converge to a single universal curve for both below and above $\rm{\textit{T}_{C}}$ for different values of magnetic field. This merging to single universal curve indicates that the compound undergo second order ferromagnetic to paramagnetic transition.

\subsection{\label{sec:level2}Summary}
In summary, we have successfully synthesized a new polycrystalline compound $\rm{Pr_2Rh_2Ga}$. This compound crystallizes in the orthorhombic, $\rm{La_2Ni_3}$-type of structure. Magnetic, heat capacity measurements and resistivity results revealed that the present compound undergoes ferromagnetic behavior with the Curie temperature $T_C$ = 18 K. We found that $\rm{Pr_2Rh_2Ga}$ does not have structural phase transition like $\mathrm{Ce_2Rh_2Ga}$. The Sommerfeld coefficient value derived from heat capacity shows, a significant enhancement, which gives an indication of a heavy electron ground state in $\rm{Pr_2Rh_2Ga}$. Resistivity results confirmed that the compound has a metallic-like conductivity character, where the curvature paramagnetic region is observed due to strong electron-phonon scattering. Below $T_C$, the temperature dependence of heat capacity and resistivity data were well described with the ferromagnetic spin-wave relation. Ferromagnetic spin-wave relation below $T_C$ yields an energy gap value of 8.47(3) K and 9.01(3) K from temperature dependent heat capacity and resistivity data, respectively. This compound exhibits both positive and negative magnetoresistance in the magnetically order state. Arrot plot ($H/M~vs.~M^{2}$) and the universal scaling plot of $-\Delta \textit{S}^{max}_{\rm M}$ $vs.$ rescaled temperature ($\theta$) confirm that this compound undergoes a second order paramagnetic to ferromagnetic phase transition. The maximum $-\Delta \textit{S}^{max}_{\rm M}$ value of $8.5~\rm J/kg.K$ and maximum $\rm \Delta \textit{T}_{ad}$ value of 3.6 K are obtained for $\rm{Pr_{2}Rh_{2}Ga}$ under change of field $0\textendash9~\rm T$. The corresponding values of RCP is 121 J/kg at 9 T. These obtained values are considerable as compare to the reported MCE materials.

\subsection*{\label{sec:level3}Acknowledgements}   
This work is supported by Global Excellence and Stature (UJ-GES) fellowship, University of Johannesburg, South Africa. AMS thanks the URC/FRC of UJ for assistance of financial support.

\subsection*{\label{ref}References}

%\nocite{*}
%\bibliography{MTJ-STO}% Produces the bibliography via BibTeX.

\end{document}